\documentclass[submission,copyright,creativecommons]{eptcs}
\providecommand{\event}{Linearity and TLLA 2018} % Name of the event you are submitting to
\usepackage{breakurl}             % Not needed if you use pdflatex only.
\usepackage{underscore}           % Only needed if you use pdflatex.
\usepackage{amsmath,amssymb,graphicx,proof}

\title{From Linear Logic to Cyclic Sharing}
\author{Masahito Hasegawa
\institute{Research Institute for Mathematical Sciences\\
Kyoto University\\%\thanks{A fine university.}\\
Kyoto, Japan}
\email{hassei@kurims.kyoto-u.ac.jp}
}
\def\titlerunning{From Linear Logic to Cyclic Sharing}
\def\authorrunning{M. Hasegawa}
\newtheorem{theorem}{Theorem}

\newtheorem{proposition}{Proposition}

\newtheorem{remark}{Remark}
\newtheorem{example}{Example}
\newcommand{\comment}[1]{}

\newcommand\Tr{\operatorname{Tr}}
\newcommand{\Int}{\mathbf{Int}}
\newcommand{\id}{\mathit{id}}
\newcommand{\sfbe}{\mathsf{be}}
\newcommand{\letbe}[3]{\mathsf{let}~#1~\mathsf{be}~#2~\mathsf{in}~#3}
\newcommand{\letrec}[2]{\mathsf{letrec}~#1~\mathsf{in}~#2}
\newcommand{\ilambda}{\lambda\!\!\!\!\!\lambda}
\newcommand{\sfC}{\mathsf{C}}
\newcommand{\sem}[1]{[\![#1]\!]}
\newcommand{\ptom}[1]{#1^+\!\Rightarrow\!#1^-}
\newcommand{\mypar}{\,\rotatebox[origin=c]{180}{\small\sf\&}\,}
\newcommand{\idline}[3]{\put(#1,#2){\line(1,0){#3}}}

\newcounter{doubleV}
\newcounter{XplusH}
\newcounter{YplusV}
\newcounter{XplusHH}
\newcounter{YplusVV}
\newcounter{symmtmp}
\newcount\mycount
                
\newcommand{\trace}[4]
    {\setcounter{doubleV}{#4}
     \addtocounter{doubleV}{#4}
     \setcounter{XplusH}{#1}
     \addtocounter{XplusH}{#3}
     \setcounter{YplusV}{#2}
     \addtocounter{YplusV}{#4}
     \put(#1,#2){\oval(\value{doubleV},\value{doubleV})[l]}
     \put(#1,\value{YplusV}){\line(1,0){#3}}
     \put(\value{XplusH},#2){\oval(\value{doubleV},\value{doubleV})[r]}
     }

\newcommand{\braid}[3]
     {\setcounter{XplusH}{#1}
      \addtocounter{XplusH}{#3}
     \setcounter{YplusV}{#2}
     \addtocounter{YplusV}{#3}
     \put(#1,#2){\line(1,1){#3}}
     \mycount=#3 \divide \mycount by 2
     \setcounter{XplusHH}{#1}
     \addtocounter{XplusHH}{\mycount}
     \addtocounter{XplusHH}{-2}
     \setcounter{YplusVV}{#2}
     \addtocounter{YplusVV}{\mycount}
     \addtocounter{YplusVV}{2}
     \qbezier(#1,\value{YplusV})(#1,\value{YplusV})(\value{XplusHH},\value{YplusVV})
     \addtocounter{XplusHH}{4}
     \addtocounter{YplusVV}{-4}
     \qbezier(\value{XplusH},#2)(\value{XplusH},#2)(\value{XplusHH},\value{YplusVV})
}

\newcommand{\braidInv}[3]
     {\setcounter{XplusH}{#1}
      \addtocounter{XplusH}{#3}
     \setcounter{YplusV}{#2}
     \addtocounter{YplusV}{#3}
     \put(#1,\value{YplusV}){\line(1,-1){#3}}
     \mycount=#3 \divide \mycount by 2
     \setcounter{XplusHH}{#1}
     \addtocounter{XplusHH}{\mycount}
     \addtocounter{XplusHH}{-2}
     \setcounter{YplusVV}{#2}
     \addtocounter{YplusVV}{\mycount}
     \addtocounter{YplusVV}{-2}
     \qbezier(#1,#2)(#1,#2)(\value{XplusHH},\value{YplusVV})
     \addtocounter{XplusHH}{4}
     \addtocounter{YplusVV}{4}
     \qbezier(\value{XplusH},\value{YplusV})(\value{XplusH},\value{YplusV})(\value{XplusHH},\value{YplusVV})
}

\begin{document}
\maketitle

\begin{abstract}
We present a translation from Multiplicative Exponential Linear Logic
to a simply-typed lambda calculus with cyclic sharing. This translation
is derived from a simple observation on the Int-construction on traced 
monoidal categories. It turns out that the translation
is a mixture of the call-by-name CPS translation and the Geometry of 
Interaction-based interpretation.
\end{abstract}

\section{Introduction}\label{sec:intro}

It is often said that {\em linear logic} \cite{Gir87} is a resource-sensitive logic.
Although this captures only one of the possible interpretations of linear logic,
it is true that we can neatly represent various resource-conscious
phenomena in linear logic. In particular, {\em sharing of
resources} can be faithfully (and fully) interpreted in linear logic:
there is a fully complete translation from the calculus of 
acyclic sharing graphs (term graphs) to the linear lambda 
calculus of the multiplicative exponential intuitionistic linear logic
(MEILL) \cite{BGHP98,Has99}.%
\footnote{%
In \cite{BGHP98} the translation from Milner's action calculi \cite{Mil96}
to MEILL is 
shown to be faithful (equationally complete), while action calculi
correspond to calculi of acyclic sharing graphs \cite{GH97,Has97}. 
Later the translation has been shown to be fully complete \cite{Has99}.}

In this paper we consider a translation in the other direction, with
suitably extended source and target calculi. Specifically, we give a
translation from a linear lambda calculus for 
multiplicative exponential linear logic (MELL)
to a simply-typed lambda calculus with cyclic sharing \cite{Has97a} 
(higher-order cyclic sharing theory \cite{Has97}).
The translation is derived from the following %simple
observation on {\em traced monoidal categories} and the {\em Int-construction}
\cite{JSV96}:
\begin{theorem}\label{thm:main}
Consider a %strong symmetric monoidal 
functor $F:\mathcal{C}\rightarrow\mathcal{D}$
from a %symmetric monoidal 
category $\mathcal{C}$ to a traced symmetric monoidal
category $\mathcal{D}$.  
Let $N:\mathcal{D}\rightarrow\Int\,\mathcal{D}$ be the 
canonical embedding of $\mathcal{D}$ to the compact closed category
$\Int\,\mathcal{D}$ obtained by the Int-construction.
Then the composition 
$\mathcal{C}\stackrel{F}{\rightarrow}\mathcal{D}\stackrel{N}{\rightarrow}\mathbf{Int}\,\mathcal{D}$
has a right adjoint if and only if
the functor 
$\mathcal{C}\stackrel{F}{\rightarrow}\mathcal{D}\stackrel{(-)\otimes D}{\longrightarrow}\mathcal{D}$
has 
a right adjoint for all $D$.
\end{theorem}
We give a proof of this theorem  later, but 
it is embarrassingly short and easy; readers
familiar with relevant concepts should have no difficulty in showing this
by themselves. 

Theorem \ref{thm:main}, when combined with the categorical semantics,
can be applied to turn a model of the lambda calculus with
cyclic sharing to a model of MELL, hence to give a translation from
MELL to the cyclic lambda calculus. Of course, the Int-construction has been
widely used to construct models of linear logic, most notably in the context
of Geometry of Interaction \cite{Gir89,AHS02}. Naturally our translation of the purely linear (multiplicative)
part is essentially the same as the standard Int- or GoI-based interpretation.
The novelty of this work lies in the treatment of 
the exponential modality $!$, which is a %direct 
consequence of Theorem \ref{thm:main}.
One small surprise is that the interpretation of intuitionistic  (or non-linear)
implication $\sigma\rightarrow\tau\,=\,\,!\sigma\multimap\tau$ agrees with
the standard {\em call-by-name continuation-passing style (CPS) translation}.
It turns out that this coincidence naturally follows from our categorical axiomatics.

We emphasize the {\em semantics-directed} nature of this work;
the construction on semantic models comes first, which in turn 
gives rise to a syntactic translation whose soundness is 
guaranteed by construction.

\paragraph{Plan of this paper}

%We proceed as follows. 
In Section \ref{sec:categories}, we recall the categorical structure 
relevant to this work and give a proof of Theorem \ref{thm:main}. 
In Section \ref{sec:models}, we recall the categorical semantics of MELL
\cite{See89,Bie95,HS03,Mel09}
and the lambda calculus with cyclic sharing \cite{Has97,Has99,PT99}, 
and see how Theorem \ref{thm:main} gives rise to
a translation from MELL to the cyclic lambda calculus.
We then recall the calculi in Section \ref{sec:calculi} 
%(and the linear lambda calculus DCLL \cite{Has05} in Appendix \ref{sec:DCLL})
and 
describe the translation concretely (Section \ref{sec:translation}). 
In Section \ref{sec:CPS}, we study the coincidence with the call-by-name 
CPS translation in terms of our model constructions.

%}
.

\section{Traced monoidal categories and Int-construction}\label{sec:categories}

\subsection{Preliminaries}
Recall that a {\em trace} \cite{JSV96} on a symmetric monoidal category
$\mathcal{C}$ is a family of maps
$$\Tr^X_{A,B}:\mathcal{C}(A\otimes X,B\otimes X)\rightarrow\mathcal{C}(A,B)$$
subject to a few axioms. For $f:A\otimes X\rightarrow B\otimes X$, 
its trace $\Tr^X_{A,B}f:A\rightarrow B$ can be drawn as a ``feedback'':
\begin{center}\unitlength=.25mm
\begin{picture}(100,50)\thicklines
\put(30,0){\framebox(40,40){$f$}}
\put(5,30){\line(1,0){25}}
\put(5,10){\line(1,0){25}}
\put(70,30){\line(1,0){25}}
\put(70,10){\line(1,0){25}}
\put(0,30){\makebox(0,0){$X$}}
\put(0,10){\makebox(0,0){$A$}}
\put(100,30){\makebox(0,0){$X$}}
\put(100,10){\makebox(0,0){$B$}}
\end{picture}
\begin{picture}(50,50)
\put(25,20){\makebox(0,0){$\mapsto$}}
\end{picture}
\begin{picture}(100,50)\thicklines
\put(30,0){\framebox(40,40){$f$}}
\trace{30}{40}{40}{10}
\put(5,10){\line(1,0){25}}
\put(70,10){\line(1,0){25}}
\put(0,10){\makebox(0,0){$A$}}
\put(100,10){\makebox(0,0){$B$}}
\end{picture}
\end{center}
A {\em traced symmetric monoidal category} is a symmetric monoidal category
equipped with a trace.%
\footnote{%
The contents of this section are valid for traced balanced monoidal categories
and tortile (ribbon) categories as well.
}%

%\subsection{Compact closed categories}
A {\em duality} 
between two objects $A$ and $B$ of
a monoidal category is determined
by a pair of arrows $\eta:I\longrightarrow B\otimes A$ and 
$\varepsilon:A\otimes B\longrightarrow I$ such that the triangle equalities
$(\varepsilon\otimes\id_A)\circ(\id_A\otimes\eta)=\id_A$ and
$(\id_B\otimes\varepsilon)\circ(\eta\otimes\id_B)=\id_B$
\comment{%%%
\begin{center}
\unitlength=.25mm
\begin{picture}(80,50)\thicklines
\put(10,20){\framebox(20,30){$\eta$}}
\put(50,0){\framebox(20,30){$\varepsilon$}}
\put(0,5){\line(1,0){50}}
\put(30,45){\line(1,0){50}}
\put(30,25){\line(1,0){20}}
\put(0,10){\makebox(0,0){$A$}}
\put(40,30){\makebox(0,0){$B$}}
\put(80,50){\makebox(0,0){$A$}}
\end{picture}
\begin{picture}(30,50)
\put(15,25){\makebox(0,0){$=$}}
\end{picture}
\begin{picture}(40,50)\thicklines
\put(0,25){\line(1,0){40}}
\put(0,30){\makebox(0,0){$A$}}
\put(40,30){\makebox(0,0){$A$}}
\end{picture}
\unitlength=.25mm
\begin{picture}(80,50)
\put(40,25){\makebox(0,0){and}}
\end{picture}
\unitlength=.25mm
\begin{picture}(80,50)\thicklines
\put(10,0){\framebox(20,30){$\eta$}}
\put(50,20){\framebox(20,30){$\varepsilon$}}
\put(0,45){\line(1,0){50}}
\put(30,5){\line(1,0){50}}
\put(30,25){\line(1,0){20}}
\put(0,50){\makebox(0,0){$B$}}
\put(40,30){\makebox(0,0){$A$}}
\put(80,10){\makebox(0,0){$B$}}
\end{picture}
\begin{picture}(30,50)
\put(15,25){\makebox(0,0){$=$}}
\end{picture}
\begin{picture}(40,50)\thicklines
\put(0,25){\line(1,0){40}}
\put(0,30){\makebox(0,0){$B$}}
\put(40,30){\makebox(0,0){$B$}}
\end{picture}
\end{center}
}%%%
hold. We say that $B$ is a {\em right dual} of $A$, and $A$ a 
{\em left dual} of $B$.
A {\em compact closed category} is a symmetric monoidal category
in which every object has a (right) dual.

%\subsection{Int-construction}

Any monoidal full subcategory of a compact closed category is traced.
Conversely, a traced symmetric monoidal category $\mathcal{C}$ gives rise to
a compact closed category $\Int\,\mathcal{C}$ to which $\mathcal{C}$
is fully faithfully embedded \cite{JSV96}.
An object of $\mathbf{Int}\,\mathcal{C}$ is a pair of objects of $\mathcal{C}$.
A morphism $f:(X,U)\rightarrow(Y,V)$ in $\mathbf{Int}\,\mathcal{C}$ 
is a morphism $ f:X\otimes V\rightarrow Y\otimes U$ in $\mathcal{C}$.
\comment{%%%
%which we shall draw as
\begin{center}%\setunitlength
\unitlength=.21mm
\begin{picture}(80,50)(0,-5)\thicklines
\put(0,-5){\dashbox{1}(80,50){}}

\put(25,0){\framebox(30,40){$f$}}
\put(0,30){{\line(1,0){25}}}
\put(0,10){{\line(1,0){25}}}
\put(55,30){{\line(1,0){25}}}
\put(55,10){{\line(1,0){25}}}

\put(-10,30){\makebox(0,0){$V$}}
\put(-10,10){\makebox(0,0){$X$}}
\put(90,30){\makebox(0,0){$U$}}
\put(90,10){\makebox(0,0){$Y$}}

\qbezier[5](0,20)(5,20)(10,20)
\qbezier[5](70,20)(75,20)(80,20)
\end{picture}
\end{center}
}%%%
The composition of 
$f:(X,U)\rightarrow(Y,V)$ and $ g:(Y,V)\rightarrow(Z,W)$
is 
\begin{center}\unitlength=0.21mm
{%\footnotesize 
\begin{picture}(200,90)(0,-10) \thicklines
\put(0,-10){\dashbox{1}(200,90){}}
\qbezier[5](0,20)(5,20)(10,20)
\qbezier[5](190,20)(195,20)(200,20)
%\put(0,0){\framebox(200,70){\mbox{}}}
 {\trace{20}{60}{160}{10}}
%\put(200,63){$-$}
%\put(20,30){\usebox{\braidd}}
\braid{20}{30}{20}
{\idline{40}{50}{50} \idline{40}{30}{10}}
\put(50,0){\framebox(30,40){\normalsize$f$}} {\idline{80}{30}{10}}
%\put(90,30){\usebox{\braiddInv}} 
\braidInv{90}{30}{20}
{\idline{110}{50}{50} \idline{110}{30}{10}}
\put(120,0){\framebox(30,40){\normalsize$g$}} {\idline{150}{30}{10}}
%\put(160,30){\usebox{\braidd}} 
\braid{160}{30}{20}
{\idline{0}{30}{20}}\put(-15,33){$W$}
{\idline{0}{10}{50}}\put(-15,-5){$X$}
{\idline{80}{10}{40}}\put(100,-5){$Y$}
{\idline{180}{30}{20}}\put(200,33){$U$}
{\idline{150}{10}{50}}\put(200,-5){$Z$}
\end{picture}}
\end{center}
The tensor product of $(X_1,U_1)$ and $(X_2,U_2)$ is $(X_1\otimes X_2,U_2\otimes U_1)$,
while the unit object is $(I,I)$.  The dual of $(X,U)$ is $(U,X)$.
There is a full faithful traced strong symmetric monoidal functor
$N:\mathcal{C}\rightarrow\mathbf{Int}\,\mathcal{C}$
sending $X$ to $(X,I)$.

\subsection{The embarrassingly easy theorem}
Now we are ready to show Theorem \ref{thm:main}:
\begin{quote}\em
Consider a %strong symmetric monoidal 
functor $F:\mathcal{C}\rightarrow\mathcal{D}$
from a %symmetric monoidal 
category $\mathcal{C}$ to a traced symmetric monoidal
category $\mathcal{D}$.  
Then 
$N\circ F:\mathcal{C}\rightarrow\Int\,\mathcal{D}$
%$\mathcal{C}\stackrel{F}{\rightarrow}\mathcal{D}\stackrel{N}{\rightarrow}\mathbf{Int}\,\mathcal{D}$
has a right adjoint if and only if
$F(-)\otimes D:\mathcal{C}\rightarrow\mathcal{D}$
%$\mathcal{C}\stackrel{F}{\rightarrow}\mathcal{D}\stackrel{(-)\otimes D}{\longrightarrow}\mathcal{D}$
has 
a right adjoint for all $D$.
\end{quote}
The proof is immediate: we have
$$
\Int \mathcal{D}(N(F(C)),(Y,D))
~=~
\Int \mathcal{D}((FC,I),(Y,D))
~\cong~
\mathcal{D}(FC\otimes D,Y)
$$%
Obviously, $N\circ F$ has a right adjoint iff 
$F(-)\otimes D:\mathcal{C}\rightarrow\mathcal{D}$
has a right adjoint for every $D$.

\begin{example}
By letting $F$ be the identity functor on $\mathcal{D}$, Theorem \ref{thm:main}
says that, for any traced symmetric monoidal category 
$\mathcal D$, $N:\mathcal{D}\rightarrow\Int\,\mathcal{D}$ has a 
right adjoint if and only if $\mathcal{D}$ is closed \cite{Has09}.
\end{example}

\begin{example}
Let $\mathbf{1}$ be a one object one arrow category.
Giving a functor from $\mathbf{1}$ to a traced symmetric monoidal category
$\mathcal{D}$ is the same as giving an object $X$ of $\mathcal{D}$. 
That $\mathbf{1}\rightarrow\mathcal{D}\stackrel{N}{\rightarrow}\Int\,\mathcal{D}$ has
a right adjoint means $N(X)=(X,I)$ is an initial object of $\Int\,\mathcal{D}$ .
Therefore, Theorem \ref{thm:main} implies that, for an object $X$ of $\mathcal{D}$, 
$(X,I)$ is an initial object of $\Int\,\mathcal{D}$ 
if and only if $X\otimes D$ is an initial object of $\mathcal{D}$ for every $D$,
i.e., $X$ is a distributive initial object in $\mathcal{D}$.
\end{example}

\section{Categorical models of linear logic and cyclic sharing}
\label{sec:models}

\subsection{Categorical models of linear logic}

A categorical model of Multiplicative Exponential Linear Logic consists of
a {\em $*$-autonomous category} $\mathcal{D}$ and
a {\em linear exponential comonad} $!$ on $\mathcal{D}$, where
\begin{itemize}
\item
A {$*$-autonomous category} \cite{Bar79}
is a symmetric monoidal closed category
equipped with an object $\bot$ such that the canonical map 
$D\longrightarrow (D\multimap\bot)\multimap\bot$ is an isomorphism for every $D$.
\item
A {linear exponential comonad} \cite{Bie95,HS03}
on a symmetric monoidal category is 
a symmetric monoidal comonad
%(comonad whose functor is symmetric monoidal and counit/comultiplication are monoidal natural)
such that its category of coalgebras is a category of commutative comonoids.
\end{itemize}
When $!$ is a linear exponential comonad, 
its category of coalgebras  is a cartesian category (the induced monoidal product is
cartesian).
Conversely, any comonad induced by 
a symmetric monoidal adjunction between a cartesian
category and a symmetric monoidal category is a linear exponential comonad
\cite{Mel09}.

\subsection{Categorical models of higher-order cyclic sharing}

A {\em Freyd category} \cite{PT99} consists of
a cartesian category %(category with finite products) 
$\mathcal{C}$,
a symmetric (pre)monoidal category $\mathcal{D}$ and
an identity-on-objects strict symmetric (pre)monoidal functor
$F:\mathcal{C}\rightarrow\mathcal{D}$.
Below we are interested in Freyd categories $F:\mathcal{C}\rightarrow\mathcal{D}$
in which $\mathcal{D}$ is monoidal (except Sec. \ref{sec:CPS}). 
% (unless otherwise stated).
A Freyd category $F:\mathcal{C}\rightarrow\mathcal{D}$ is 
\begin{itemize}
\item {\em closed} when the functor $F(-)\otimes D$ has a right adjoint 
(called {\em Kleisli exponential}) 
$D\Rightarrow(-)$ for each $D$. 
\item {\em traced} when $\mathcal{D}$ is traced.
\end{itemize}
We employ the approach taken in our previous work of modelling
sharing graphs \cite{Has97}
using Freyd categories as the key semantic structure.
\begin{enumerate}
\item
For modelling {\em first-order acyclic sharing}, we use a 
{\em Freyd category}
$F:\mathcal{C}\rightarrow\mathcal{D}$
where $\mathcal{D}$ is supposed to be monoidal.  
Values (including variables=wires) are interpreted in the
cartesian category $\mathcal{C}$, while terms with sharing are
interpreted in the monoidal category $\mathcal{D}$. 
\item
For modelling {\em higher-order structure} (i.e. lambda abstraction and
application), 
we use a {\em  closed Freyd category}.  The adjunction
$\mathcal{D}(FC\otimes D,E)\cong\mathcal{C}(C,D\Rightarrow E)$
says that currying turns a term to a value.
\item
For modelling {\em cyclic sharing}, we use a 
{\em traced Freyd category}.
\item
Finally, for modelling {\em higher-order cyclic sharing (lambda calculus with cyclic sharing)}, 
we use a {\em traced closed Freyd category}.
\end{enumerate}

\subsection{Relating models}

We apply Theorem \ref{thm:main} to turn a 
model of higher-order cyclic sharing (traced closed Freyd category)
to a model of MELL ($*$-autonomous category with a linear exponential
comonad).
As an immediate corollary to the theorem, we have:
\begin{proposition}
A traced Freyd category
$F:\mathcal{C}\rightarrow\mathcal{D}$
is closed if and only if $N\circ F:\mathcal{C}\rightarrow\Int\,\mathcal{D}$
has a right adjoint.
\end{proposition}
Now suppose that  we have a traced closed Freyd category 
$F:\mathcal{C}\rightarrow\mathcal{D}$ with
$ F(-)\otimes D\dashv D\Rightarrow(-)$ for each $D$.
Then the strong symmetric monoidal functor 
$NF:\mathcal{C}\rightarrow\Int \mathcal{D}$ has
a right adjoint sending $(X,U)$ to $U\Rightarrow X$.
This symmetric monoidal adjunction gives rise to 
a linear exponential comonad $!$
on $\Int\,\mathcal{D}$
sending $(X,U)$ to $(U\Rightarrow X,I)$.
Since $\Int\,\mathcal{D}$ is compact closed, it is
$*$-autonomous. 
Thus $\Int\,\mathcal{D}$ with $!$ 
gives a model of MELL.

By applying this construction to the term model of the
lambda calculus with cyclic sharing, we obtain a translation 
from MELL to the cyclic lambda calculus. This will be spelled
out in the rest of this paper.

\section{The calculi} \label{sec:calculi}

\subsection{A lambda calculus with cyclic sharing}

We give a simply typed lambda calculus with cyclic sharing $\lambda_\mathsf{letrec}$
(Figure \ref{fig:letrec}). This is essentially the same calculus as the 
higher-order cyclic sharing theory in \cite{Has99}, but slightly modified for
a better match with semantic models (closed traced Freyd categories);
the only differences are (i) treatment of product types (strictly associative or not)
and (ii) treatment of variables (allowing variables on product types or not).
Readers familar with Moggi's computational lambda calculus \cite{Mog89}
should note that
the $\lambda_\mathsf{letrec}$-calculus can be regarded as the
commutative\footnote{%
Here ``commutativity'' means the commutativity of effects
$
\letbe{x}{L}{\letbe{y}{M}{N}}~=~
\letbe{y}{M}{\letbe{x}{L}{N}}
$.
At the level of semantic models, this amounts to forcing the Freyd categories
to be monoidal (rather than premonoidal), and to assuming the strong monads
to be commutative.
}
version of the computational lambda calculus
enriched with the recursive let-binding {\sf letrec} for 
expressing cyclic sharing, as emphasized in \cite{Has97a}.

As shown in \cite{Has99}, the $\lambda_\mathsf{letrec}$-calculus is 
sound and complete for models given by closed traced Freyd categories.

\begin{figure}\small
\paragraph{Types}
$
\sigma~::=~b~|~\sigma\Rightarrow\sigma~|~\sigma\times\sigma~|~1
$

\mbox{}

\paragraph{Declarations, terms, values and contexts}
$$
\begin{array}{llcl}
\mbox{Declarations}&
D &::=& \emptyset~|~x^\sigma~\sfbe~M,D\\
\mbox{Terms} &
M &::=& x~|~\lambda x^\sigma.M~|~M\,M~|~(M,M)~|~\pi_i\,M~|~*~|~\letrec{D}{M}
\\
\mbox{Values}&
V &::=& x~|~\lambda x^\sigma.M~|~(V,V)~|~\pi_i\,V~|~*
\\
\mbox{Contexts}&
C &::=& [-]~|~C\,M~|~M\,C~|~(C,M)~|~(M,C)
\end{array}
$$
In the term $\letrec{D}{M}$, the declaration $D$ must be non-empty.

\paragraph{Typing}

$$
\begin{array}{ccc}
\infer
{\Gamma_1,x:\sigma,\Gamma_2\vdash x:\sigma}
{}
&
\infer
{\Gamma\vdash\lambda x^\sigma.M:\sigma\Rightarrow\tau}
{\Gamma,x:\sigma\vdash M:\tau}
&
\infer
{\Gamma\vdash M\,N:\tau}
{\Gamma\vdash M:\sigma\Rightarrow\tau & \Gamma\vdash N:\sigma}
\\
\\
\infer
{\Gamma\vdash (M,N):\sigma\times\tau}
{\Gamma\vdash M:\sigma & \Gamma\vdash N:\tau}
&
\infer
{\Gamma\vdash \pi_i\,M:\sigma_i}
{\Gamma\vdash M:\sigma_1\times\sigma_2}
&
\infer
{\Gamma\vdash *:1}
{}
\end{array}
$$
$$
\infer
{\Gamma\vdash\letrec{x_1^{\sigma_1}~\sfbe~M_1,\dots,x_n^{\sigma_n}~\sfbe~M_n}{N}:\sigma}
{\Gamma,x_1:\sigma_1,\dots,x_n:\sigma_n\vdash M_i:\sigma_i~(i=1,\dots,n)
 &
 \Gamma,x_1:\sigma_1,\dots,x_n:\sigma_n\vdash N:\sigma}
$$

\paragraph{Notations}
We make use of the following syntax sugar.
$$
\begin{array}{rcl}
\letbe{x^\sigma}{M}{N} &\equiv& (\lambda x^\sigma.N)\,M\\
\lambda(x^{\sigma_1},y^{\sigma_2}).M &\equiv&
 \lambda z^{\sigma_1\times\sigma_2}.M[x:=\pi_1\,z,y:=\pi_2\,z]\\
\letbe{(x^{\sigma_1},y^{\sigma_2})}{M}{N} &\equiv &
 (\lambda(x^{\sigma_1},y^{\sigma_2}).N)\,M\\
\letrec{D_1,(x,y)~\sfbe~M,D_2}{N} &\equiv&
 \letrec{D_1,z~\sfbe~M,x~\sfbe~\pi_1\,z,y~\sfbe~\pi_2\,z,D_2}{N}
\end{array}
$$

\paragraph{Axioms}
$$
\begin{array}{lrcll}
\beta^\Rightarrow_v &
(\lambda x.M)\,V &=& M[x:=V]\\
\eta^\Rightarrow_v &
\lambda x.V\,x &=& V ~~~~~ (x\not\in FV(V))\\
\beta^\times_v &
\pi_i\,(V_1,V_2) &=& V_i\\
\eta^\times_v &
(\pi_1\,V,\pi_2\,V) &=& V & \\
\beta^1_v &
V &=& * ~~~~~ (V:1)\\
\mathit{Comm}_\mathsf{let} &
C[M] &=& \letbe{x}{M}{C[x]}~~~~~ (x\not\in FV(C)) \\
\\
%\end{array}
%$$
%$$
%\begin{array}{lrcll}
\mathit{Assoc}_1 &
\letrec{x~\sfbe~(\letrec{D_1}{M}),D_2}{N}
&=&
\multicolumn{2}{l}{\letrec{D_1,x~\sfbe~M,D_2}{N}}\\
\mathit{Assoc}_2 &
\letrec{D_1}{\letrec{D_2}{M}} &=& \letrec{D_1,D_2}{M}\\
\mathit{Perm} &
\letrec{D_1,x~\sfbe~L,y~\sfbe~M,D_2}{N} &=& 
\multicolumn{2}{l}{\letrec{D_1,y~\sfbe~M,x~\sfbe~L,D_2}{N}} \\
\mathit{Comm}_\mathsf{letrec} &
C[\letrec{D}{M}] &=& \letrec{D}{C[M]}\\
\sigma_1&
\letrec{x~\sfbe~V,D}{M}
&=&
\letrec{x~\sfbe~V,D[x:=V]}{M[x:=V]}
\\
\sigma_2 &
\letrec{x~\sfbe~M}{N} &=& \letbe{x}{M}{N} ~~~~~~(x\not\in FV(M))\\
%\sigma_2 &
%\letrec{x~\sfbe~V,D[x]}{M} &=& \letrec{x~\sfbe~V,D[V]}{M}\\
%\sigma_3 &
%\letrec{x~\sfbe~V,D}{M[x]} &=& \letrec{x~\sfbe~V,D}{M[V]}\\
\end{array}
$$
\caption{The $\lambda_\mathsf{letrec}$-calculus} 
\label{fig:letrec}
\end{figure}

\subsection{A linear lambda calculus for MELL}

As the calculus for MELL, we use DCLL (dual classical linear logic) \cite{Has05},
as recalled in Figure \ref{fig:DCLL}. %Appendix \ref{sec:DCLL}.
DCLL is an extension of DILL (dual intuitionistic linear logic) of Barber and Plotkin
\cite{BP97}, but has just linear implication $\multimap$,
non-linear implication $\rightarrow$ and the falsity type $\bot$ as the primitive type
constructs. Terms are built from variables,
the linear lambda abstraction $\lambda x^\sigma.M$ and application
$M\,N$, non-linear lambda abstraction $\ilambda x^\sigma.M$ and
application $M@ N$, 
and the double-negation elimination $\sfC_\sigma\,M$.
%lambda abstraction/applicaton (for both linear and non-linear implications) 
%and a combinator for double-negation elimination.  
Like DILL, DCLL employs a dual-context formulation,
where a typing judgement takes the form $\Gamma~;~\Delta\vdash M:\tau$
in which $\Gamma$ represents a non-linear (intuitionistic) context 
whose variables can be used as many times as we like whereas
$\Delta$ is a linear context whose variables are used exactly once.
The equational theory has just the 
$\beta\eta$-axioms together with two axioms for the isomorphism
$ (\sigma\multimap\bot)\multimap\bot\,\cong\,\sigma$.

Despite its simplicity, DCLL is sound and complete for models given by
$*$-autonomous categories with a linear exponential comonad,
and can express other connectives and proofs of MELL  \cite{Has05},
e.g., 
$I=\bot\multimap\bot$, 
$\sigma\otimes\tau=(\sigma\multimap\tau\multimap\bot)\multimap\bot$
and
$!\sigma=(\sigma\rightarrow\bot)\multimap\bot$.

\begin{figure}\small

\paragraph{Types and terms}

$$
\begin{array}{rcl}
\sigma &::=& b~|~\bot~|~\sigma\multimap\sigma~|~\sigma\rightarrow\sigma
\\
M  &::=&  x~|~\lambda x^\sigma.M~|~M\,M~|~\ilambda x^\sigma.M~|~M@M~|~\sfC_\sigma\,M
\end{array}
$$
where $b$ ranges over a set of base types. We may omit the type subscripts for
ease of presentation.

\paragraph{Typing}

$$
\begin{array}{cc}
\infer[\mathit{Ax}_\mathit{nonlinear}]
{\Gamma_1,x:\sigma,\Gamma_2~;~\emptyset\vdash x:\sigma}
{}
&
\infer[\mathit{Ax}_\mathit{linear}]
{\Gamma~;~x:\sigma\vdash x:\sigma}
{}
\\
\\
\infer[\multimap\mathit{Intr}]
{\Gamma~;~\Delta\vdash\lambda x^\sigma.M:\sigma\multimap\tau}
{\Gamma~;~\Delta,x:\sigma\vdash M:\tau}
&
\infer[\multimap\mathit{Elim}]
{\Gamma~;~\Delta_1\sharp\Delta_2\vdash M\,N:\tau}
{\Gamma~;~\Delta_1\vdash M:\sigma\multimap\tau 
 &
 \Gamma~;~\Delta_2\vdash N:\sigma}
\\
\\
\infer[\rightarrow\mathit{Intr}]
{\Gamma~;~\Delta\vdash \ilambda x^\sigma.M:\sigma\rightarrow\tau}
{\Gamma,x:\sigma~;~\Delta\vdash M:\tau}
&
\infer[\rightarrow\mathit{Elim}]
{\Gamma~;~\Delta\vdash M@N:\tau}
{\Gamma~;~\Delta\vdash M:\sigma\rightarrow\tau 
 &
 \Gamma~;~\emptyset\vdash N:\sigma}
\\
\\
\multicolumn{2}{c}{
\infer[\mathit{Duality}]
{\Gamma~;~\Delta\vdash \sfC_\sigma\,M:\sigma}
{\Gamma~;~\Delta\vdash M:(\sigma\multimap\bot)\multimap\bot}
}
\end{array}
$$
where $\Delta_1\sharp\Delta_2$ is a merge of $\Delta_1$ and $\Delta_2$ \cite{BP97}.
When we introduce $\Delta_1\sharp\Delta_2$, it is assumed that 
there is no variable occurring both in $\Delta_1$ and $\Delta_2$.

\paragraph{Axioms}

$$
\begin{array}{lrcll}
\beta^\multimap &
(\lambda x.M)\,N &=& M[x:=N]\\
\eta^\multimap &
\lambda x.M\,x &=& M\\
\beta^\rightarrow &
(\ilambda x.M)@N &=& M[x:=N]\\
\eta^\rightarrow &
\ilambda x.M@x &=& M & (x\not\in FV(M))\\
\sfC_1 &
L\,(\sfC_\sigma\,M) &=& M\,L & (L:\sigma\multimap\bot)\\
\sfC_2 &
\sfC_\sigma\,(\lambda k.k\,M) &=& M\\
\end{array}
$$

\caption{Dual Classical Linear Logic (DCLL)} 
\label{fig:DCLL}
\end{figure}

\section{The translation} \label{sec:translation}

\subsection{From DCLL to the $\lambda_\mathsf{letrec}$-calculus}

We spell out the translation from DCLL to the 
$\lambda_\mathsf{letrec}$-calculus derived from the semantic
construction in Section \ref{sec:models}. 
For each base type $b$ of DCLL,  fix types $b^+$ and $b^-$ of the 
$\lambda_\mathsf{letrec}$-calculus.
For a typing judgement 
$$
\Gamma~;~\Delta  \vdash M:\sigma
$$
in DCLL, its translation in the $\lambda_\mathsf{letrec}$-calculus
$$
\Gamma^{-\Rightarrow +},\Delta^+
 \vdash \sem{M}^\Delta:\sigma^-\Rightarrow(\sigma^+\times \Delta^-)
$$
is given as follows, where 
$$
\begin{array}{rcl}
(x_1:\sigma_1,\dots,x_m:\sigma_m)^{-\Rightarrow +}
&=&
x_1:\ptom{\sigma_1},\dots, x_m:\ptom{\sigma_m}
\\
(y_1:\tau_1,\dots,y_n:\tau_n)^+
&=&
y_1:\tau_1^+,\dots,y_n:\tau_n^+
\\
(y_1:\tau_1,\dots,y_n:\tau_n)^-
&=&
\tau_1^-\times\cdots\times\tau_n^-
\end{array}
$$
and the translation of types and terms are inductively given as follows.
Figure \ref{fig:translation} gives a summary of the translation.

\begin{remark}
In describing the translation, we pretend as if the product types 
in the $\lambda_\mathsf{letrec}$-calculus are strictly associative,
e.g., we identify $((x,y),z):(\sigma_1\times\sigma_2)\times\sigma_3$ with $(x,(y,z)):\sigma_1\times(\sigma_2\times\sigma_3)$ and $(x,*):\sigma\times 1$ with $x:\sigma$.
This %allows us to replace the isomorphisms above by the identities, and 
makes the description of translation much simpler. Alternatively, we could 
make use of the original higher-order cyclic sharing theory \cite{Has97} whose 
products are strictly associative. 
\end{remark}

\begin{figure}[t]
$$\small
\renewcommand{\arraystretch}{.5}
\begin{array}{c|c}
\\
\Gamma~;~\Delta\vdash_\mathbf{DCLL} M:\sigma
&
\Gamma^{-\Rightarrow +},\Delta^+\vdash_{\lambda_\mathsf{letrec}}\sem{M}:\sigma^-\Rightarrow(\sigma^+\times\Delta^-)
\\
\\ \hline\hline
\\
\infer[\mathit{Ax}_\mathit{non\,linear}]
{\Gamma_1,x:\sigma,\Gamma_2~;~\emptyset\vdash x:\sigma}
{}
&
\infer
{\Gamma_1^{-\Rightarrow +},x:\sigma^-\Rightarrow\sigma^+,\Gamma_2^{-\Rightarrow +}\vdash \lambda k^{\sigma^-}.x\,k:\sigma^-\Rightarrow\sigma^+}
{}
\\
\\ \hline
\\
\infer[\mathit{Ax}_\mathit{linear}]
{\Gamma~;~x:\sigma\vdash x:\sigma}
{}
&
\infer
{\Gamma^{-\Rightarrow +},x:\sigma^+\vdash\lambda k^{\sigma^-}.(x,k)
:\sigma^-\Rightarrow(\sigma^+\times\sigma^-)}
{}
\\
\\ \hline
\\
\infer[\multimap\mathit{Intr}]
{\Gamma~;~\Delta\vdash\lambda x^\sigma.M:\sigma\multimap\tau}
{\Gamma~;~\Delta,x:\sigma\vdash M:\tau}
&
\infer
{\Gamma^{-\Rightarrow +},\Delta^+\vdash \lambda(x,k).\sem{M}^{\Delta,x:\sigma}\,k
:(\sigma^+\times\tau^-)\Rightarrow(\tau^+\times\Delta^-)}
{\Gamma^{-\Rightarrow +},\Delta^+,x:\sigma^+\vdash \sem{M}^{\Delta,x:\sigma}
:\tau^-\Rightarrow(\tau^+\times\Delta^-)}
\\
\\ \hline
\\
\infer[\multimap\!\mathit{Elim}]
{\Gamma~;~\Delta_1\sharp\Delta_2\vdash M\,N:\tau}
{\Gamma~;~\Delta_1\vdash M:\sigma\multimap\tau 
 &
 \Gamma~;~\Delta_2\vdash N:\sigma}
&
%\mbox{(see below)}
\infer
{
\begin{array}{l}
\Gamma^{-\Rightarrow +},
(\Delta_1\sharp\Delta_2)^+
\vdash\\
~~~~
%% corrected 30 May 2018
%%\lambda k^{\sigma_2^-}.
\lambda k^{\tau^-}.
\mathsf{letrec}~
(u:\sigma^+,
\vec{z_2}:\Delta_2^-)
~\sfbe~
\sem{N}^{\Delta_2}\,h,\\
~~~~~~~~~~~~~~~~~~~\,
(v:\tau^+,
h:\sigma^-,
\vec{z_1}:\Delta_1^-)
~\sfbe~
\sem{M}^{\Delta_1}(u,k)\\
~~~~~~~~~~~~~~~~
\mathsf{in}~
(v,\vec{z_1}\sharp\vec{z_2})
~:~
\tau^-\Rightarrow(\tau^+\times(\Delta_1\sharp\Delta_2)^-)
\end{array}
}
{
\begin{array}{c}
\Gamma^{-\Rightarrow +},\Delta_1^+\vdash
\sem{M}^{\Delta_1}:
(\sigma^+\times\tau^-)\Rightarrow
(\tau^+\times\sigma^-\times\Delta_1^-)
\\
\Gamma^{-\Rightarrow +},\Delta_2^+\vdash
\sem{N}^{\Delta_2}:
\sigma^-\Rightarrow
(\sigma^+\times\Delta_2^-)
\end{array}
}
\\
\\ \hline
\\
\infer[\rightarrow\mathit{Intr}]
{\Gamma~;~\Delta\vdash \ilambda x^\sigma.M:\sigma\rightarrow\tau}
{\Gamma,x:\sigma~;~\Delta\vdash M:\tau}
&
\infer
{\Gamma^{-\Rightarrow +},\Delta^+\vdash\lambda(x,k).\sem{M}^\Delta\,k:
 ((\sigma^-\Rightarrow\sigma^+)\times\tau^-)\Rightarrow(\tau^+\times\Delta^-)}
{\Gamma^{-\Rightarrow +},x:\sigma^-\Rightarrow\sigma^+,\Delta^+
 \vdash \sem{M}^\Delta :\tau^-\Rightarrow(\tau^+\times\Delta^-)}
\\
\\ \hline
\\
\infer[\rightarrow\mathit{Elim}]
{\Gamma~;~\Delta\vdash M@N:\tau}
{\Gamma~;~\Delta\vdash M:\sigma\rightarrow\tau 
 &
 \Gamma~;~\emptyset\vdash N:\sigma}
&
\infer
{\Gamma^{-\Rightarrow +},\Delta^+\vdash \lambda k.\sem{M}^\Delta\,(\sem{N}^\emptyset,k)
:\tau^-\Rightarrow(\tau^+\times\Delta^-)}
{
\begin{array}{c}
\Gamma^{-\Rightarrow +},\Delta^+\vdash\sem{M}^\Delta
:((\sigma^-\Rightarrow\sigma^+)\times\tau^-)\Rightarrow(\tau^+\times\Delta^-)
\\
\Gamma^{-\Rightarrow +}\vdash\sem{N}^\emptyset:\sigma^-\Rightarrow\sigma^+
\end{array}
}
\\
\\ \hline
\\
\infer[\mathit{Duality}]
{\Gamma~;~\Delta\vdash \sfC_\sigma\,M:\sigma}
{\Gamma~;~\Delta\vdash M:(\sigma\multimap\bot)\multimap\bot}
&
\comment{
\infer
{\Gamma^{-\Rightarrow +},\Delta^+\vdash
 \lambda k.\letbe{(u^1,x,v^1,\vec{z})}{\sem{M}\,(*,k,*)}{(x,\vec{z})}
:\sigma^-\Rightarrow(\sigma^+\times\Delta^-)
}
{\Gamma^{-\Rightarrow +},\Delta^+\vdash\sem{M}
:(1\times\sigma^-\times1)\Rightarrow(1\times\sigma^+\times1\times\Delta^-)}
}
\infer
{\Gamma^{-\Rightarrow +},\Delta^+\vdash\sem{M}^\Delta
:\sigma^-\Rightarrow(\sigma^+\times\Delta^-)
}
{\Gamma^{-\Rightarrow +},\Delta^+\vdash\sem{M}^\Delta
:\sigma^-\Rightarrow(\sigma^+\times\Delta^-)}
\\
\\
\end{array}
$$
\caption{Summary of the translation}
\label{fig:translation}
\end{figure}

\paragraph{Translation of types}
$$
\begin{array}{rclrcl}
(\sigma\rightarrow\tau)^+&=&\tau^+ &
(\sigma\rightarrow\tau)^-&=&(\sigma^-\Rightarrow\sigma^+)\times\tau^- \\
(\sigma\multimap\tau)^+&=&\tau^+\times\sigma^- ~~~&
(\sigma\multimap\tau)^-&=&\sigma^+\times\tau^- \\
\bot^+ &=& 1 &
\bot^- &=& 1
\end{array}
$$
%Note that 
%$((\sigma\multimap\bot)\multimap\bot)^+\cong\sigma^+$ and
%$((\sigma\multimap\bot)\multimap\bot)^-\cong\sigma^-$ hold.
\comment{%%%
For other connectives, we have
$$
\begin{array}{rclrcl}
(\sigma\otimes\tau)^+&\cong&\sigma^+\times\tau^+ &
(\sigma\otimes\tau)^-&\cong&\tau^-\times\sigma^-\\
(\sigma\mypar\tau)^+&\cong&\sigma^+\times\tau^+ &
(\sigma\mypar\tau)^-&\cong&\tau^-\times\sigma^-\\
I^+ &\cong& 1 & I^- &\cong& 1\\
(!\sigma)^+ &\cong& \sigma^-\Rightarrow\sigma^+ &
(!\sigma)^- &\cong& 1\\
(?\sigma)^+ &\cong& 1 &
(?\sigma)^- &\cong& \sigma^+\Rightarrow\sigma^-\\
\end{array}
$$
}%%%

\paragraph{Translation of terms}

$$
\begin{array}{rcll}
\sem{x}^\emptyset &=& \lambda k.x\,k \\%~~~(x\not\in\Delta)\\
\sem{y}^{y:\sigma} &=& \lambda k.(y,k)  \\% ~~~(y\in\Delta)\\
\sem{\ilambda x.M}^\Delta &=& \lambda(x,k).\sem{M}^\Delta\,k\\
\sem{M @ N}^\Delta &=& \lambda k.\sem{M}^\Delta\,(\sem{N}^\emptyset, k)\\
\sem{\lambda y^\sigma.M}^\Delta &=&  \lambda(y,k).\sem{M}^{\Delta,y:\sigma}k \\ 
\sem{M\,N}^{\Delta_1\sharp\Delta_2} &=& 
%%
%% corrected 30 May 2018
 \lambda k.(\letrec{(u,\vec{z_2})~\sfbe~\sem{N}^{\Delta_2}\,h,~
                                (v,h,\vec{z_1})~\sfbe~\sem{M}^{\Delta_1}(u,k)}
                              {(v,\vec{z_1}\sharp\vec{z_2})})
% \lambda k.(\letrec{(u,\vec{z_1})~\sfbe~\sem{N}^{\Delta_2}\,h,~
%                               (v,h,\vec{z_2})~\sfbe~\sem{M}^{\Delta_1}(u,k)}
%                             {(v,\vec{z_1}\sharp\vec{z_2})})
\\
\sem{\mathsf{C}\,M}^\Delta &=& \sem{M}^\Delta
\end{array}
$$
Note that the translation of linear constructs agrees with the standard 
Int- or GoI-based interpretation. For instance, the linear
application $\sem{M\,N}^{\Delta_1,\Delta_2}$ is graphically presented as follows -- it 
is an instance of the composition in compact closed categories
obtained by the Int-construction.
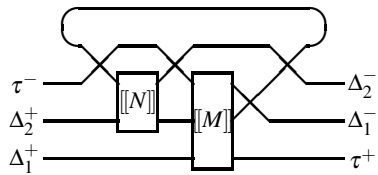
\begin{center}\unitlength=.25mm\footnotesize
\begin{picture}(160,80)(0,20)\thicklines
\put(40,35){\framebox(20,30){$\sem{N}$}}
\put(80,15){\framebox(20,50){$\sem{M}$}}
\trace{20}{90}{120}{10}
\put(0,60){\line(1,0){20}}
\put(0,40){\line(1,0){40}}
\put(0,20){\line(1,0){80}}
\put(20,80){\line(1,-1){20}}
\put(20,60){\line(1,1){20}}
\put(40,80){\line(1,0){20}}
\put(60,80){\line(1,-1){20}}
\put(60,60){\line(1,1){20}}
\put(60,40){\line(1,0){20}}
\put(80,80){\line(1,0){40}}
\put(100,60){\line(1,-1){20}}
\put(100,40){\line(1,1){40}}
\put(100,20){\line(1,0){60}}
\put(120,80){\line(1,-1){20}}
\put(120,40){\line(1,0){40}}
\put(140,60){\line(1,0){20}}
\put(-10,60){\makebox(0,0){$\tau^-$}}
\put(-10,40){\makebox(0,0){$\Delta_2^+$}}
\put(-10,20){\makebox(0,0){$\Delta_1^+$}}
\put(170,60){\makebox(0,0){$\Delta_2^-$}}
\put(170,40){\makebox(0,0){$\Delta_1^-$}}
\put(170,20){\makebox(0,0){$\tau^+$}}
\end{picture}
\end{center}

The soundness of the translation follows by definition:

\begin{proposition}[type soundness]
If the typing judgement 
$\Gamma~;~\Delta\vdash M:\sigma$ is derivable in DCLL, then 
$\Gamma^{-\Rightarrow +},\Delta^+
 \vdash \sem{M}^\Delta:\sigma^-\Rightarrow(\sigma^+\times \Delta^-)$
is derivable 
in the $\lambda_\mathsf{letrec}$-calculus.
\end{proposition}

\begin{proposition}[equational soundness]
If the equation $\Gamma~;~\Delta\vdash M=N:\sigma$ is derivable in DCLL, then $\Gamma^{-\Rightarrow +},\Delta^+
 \vdash \sem{M}=\sem{N}$ is derivable in the $\lambda_\mathsf{letrec}$-calculus.
\end{proposition}

We shall note that our translation is {\em not} equationally complete,
because of the coherence of compact closed categories.
For instance, the two proofs of 
$((\sigma\multimap I)\multimap I)\multimap I\vdash
((\sigma\multimap I)\multimap I)\multimap I$
(where $I=\bot\multimap\bot$) get the same interpretation
(the triple-unit problem \cite{KM71}).

\subsection{Examples}
One might expect that the translation of a well-typed 
term of DCLL would be equal to a $\mathsf{letrec}$-free term
in the $\lambda_\mathsf{letrec}$-calculus.
The following example shows that it is not the case;
we cannot eliminate $\mathsf{letrec}$
even when we restrict our attention to
terms in $\beta$-normal form.
Consider the term 
$$
f:\sigma\multimap\tau,~g:\tau\multimap\delta~;~x:\sigma
\vdash_\mathbf{DCLL}
g\,(f\,x):\delta
$$
with non-linear variables $f,g$ and a linear variable $x$.
This term is interpreted as
$$
\begin{array}{c}
f:(\sigma^+\times\tau^-)\Rightarrow(\tau^+\times\sigma^-),\\
g:(\tau^+\times\delta^-)\Rightarrow(\delta^+\times\tau^-),\\
x:\sigma^+
\end{array}
\vdash
\lambda k^{\delta^-}.
\Big(\letrec{
\begin{array}{c}
(u^{\tau^+},z^{\sigma^-})~\sfbe~f\,(x,h),\\
(v^{\delta^+},h^{\tau^-})~\sfbe~g\,(u,k)
\end{array}
}{(v,z)}\Big)
:\delta^-\Rightarrow(\delta^+\times\sigma^-)
$$
This is precisely the composition in the Int-category.
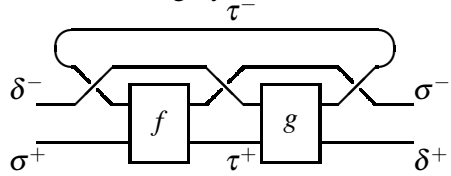
\begin{center}\unitlength=0.25mm
{%\footnotesize 
\begin{picture}(200,90)(0,-10) \thicklines
%\put(0,-10){\dashbox{1}(200,90){}}
%\qbezier[5](0,20)(5,20)(10,20)
%\qbezier[5](190,20)(195,20)(200,20)
%\put(0,0){\framebox(200,70){\mbox{}}}
 {\trace{20}{60}{160}{10}}
%\put(200,63){$-$}
%\put(20,30){\usebox{\braidd}}
\braid{20}{30}{20}
{\idline{40}{50}{50} \idline{40}{30}{10}}
\put(50,0){\framebox(30,40){\normalsize$f$}} {\idline{80}{30}{10}}
%\put(90,30){\usebox{\braiddInv}} 
\braidInv{90}{30}{20}
{\idline{110}{50}{50} \idline{110}{30}{10}}
\put(120,0){\framebox(30,40){\normalsize$g$}} {\idline{150}{30}{10}}
%\put(160,30){\usebox{\braidd}} 
\braid{160}{30}{20}
{\idline{0}{30}{20}}\put(-15,33){$\delta^-$}
{\idline{0}{10}{50}}\put(-15,-5){$\sigma^+$}
{\idline{80}{10}{40}}\put(100,-5){$\tau^+$}
{\idline{180}{30}{20}}\put(200,33){$\sigma^-$}
{\idline{150}{10}{50}}\put(200,-5){$\delta^+$}
\put(100,75){$\tau^-$}
\end{picture}}
\end{center}
Note that the use of non-linear variables is essential. 
For instance, the interpretation of the following term
$$
\emptyset~;~
f:\sigma\multimap\tau,~g:\tau\multimap\delta~,~x:\sigma
\vdash_\mathbf{DCLL}
g\,(f\,x):\delta
$$
with linear variables $f,g,x$ is equal to the $\mathsf{letrec}$-free term
$$
f:\tau^+\times\sigma^-,
g:\delta^+\times\tau^-,
x:\sigma^+
\vdash
\lambda k^{\delta^-}.(\pi_1\,g,x,\pi_2\,g,\pi_1\,f,k,\pi_2\,f)
:\delta^-\Rightarrow(\delta^+\times\sigma^+\times\tau^-\times\tau^+\times\delta^-\times\sigma^-)
$$
since the linear variables are simply interpreted as wirings.

\section{Relation to the call-by-name CPS translation}\label{sec:CPS}

The translation of non-linear variables, non-linear lambda abstraction
and non-linear application agrees with the (Streicher-style)
{\em call-by-name CPS translation}
\cite{SR98}
$$\sem{x}^\emptyset=\lambda k.x\,k~~~~~\sem{\ilambda x.M}^\emptyset=\lambda(x,k).\sem{M}^\emptyset k
~~~~~\sem{M@ N}^\emptyset=\lambda k.\sem{M}^\emptyset(\sem{N}^\emptyset,k)$$%
though our translation does not assume a single fixed answer type.
This is because the translation of non-linear types picks up a
cartesian closed category derived from the closed Freyd category:

\begin{proposition}[folklore?]
Suppose that $F:\mathcal{C}\rightarrow\mathcal{D}$ is a 
closed Freyd category %with $F(-)\otimes D\dashv D\Rightarrow(-)$
(where $\mathcal{D}$ can be premonoidal).
Then there is a cartesian closed full subcategory of $\mathcal{C}$ whose 
objects are finite products of objects of the form $D\Rightarrow E$.
\end{proposition}
$$\mathcal{C}(\Gamma\times(D\Rightarrow E),\,D'\Rightarrow E')
~\cong~
\mathcal{C}(\Gamma,\,((D\Rightarrow E)\times D')\Rightarrow E')
$$
In the case of traced closed Freyd categories, this cartesian closed category is 
%equivalent to the %cartesian closed 
%subcategory of finite products of cofree coalgebras in $(\Int\,\mathcal{D})^!$
%where $!(X,U)=(U\Rightarrow X,1)$:
where the interpretation
of non-linear abstraction and application takes place:
$$
\Int\,\mathcal{D}(!(X_1,U_1)\otimes\dots\otimes !(X_n,U_n),(Y,V))
~\cong~
\mathcal{C}((U_1\Rightarrow X_1)\times\dots\times (U_n\Rightarrow X_n),
V\Rightarrow Y)
$$
%This is where the interpretation
%of non-linear abstraction and application takes place.
By letting the codomains $E$ and $E'$ (or $X_i$ and $Y$)
be a fixed answer type, we obtain the standard
CPS semantics.

\mbox{}

This situation can be summarized as the following picture. The
outer square expresses our model constructions from 
traced closed Freyd categories, while the 
inner triangle shows the induced syntactic translations.
Commutativity at the level of semantic model consructions 
guarantees commutativity of syntactic translations.
(Since we 
employ DCLL as the language for MELL, the Girard translation from 
the simply typed lambda calculus to MELL is just an inclusion in our 
formulation.)

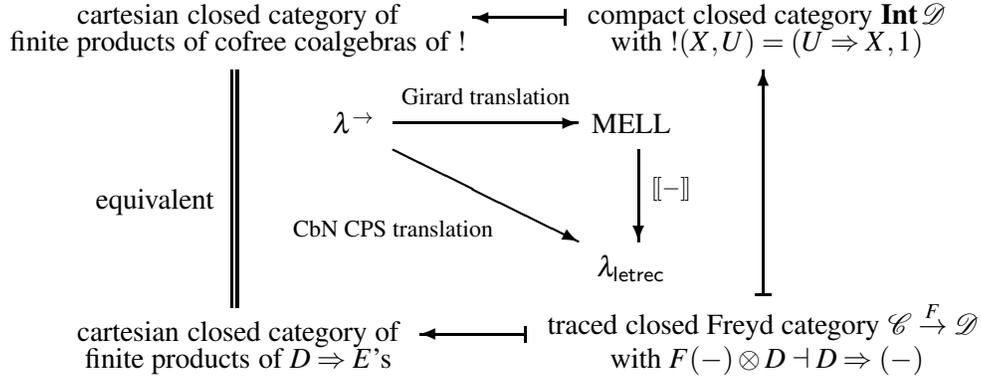
\begin{center}
\begin{picture}(200,145)(-20,-15)\thicklines
\put(130,80){\makebox(0,0){MELL}}
\put(130,70){ \vector(0,-1){35}}
\put(145,55){\makebox(0,0){\footnotesize $\sem{-}$}}
\put(130,25){\makebox(0,0){$\lambda_\mathsf{letrec}$}}

\put(25,80){\makebox(0,0){$\lambda^\rightarrow$}}
\put(40,80){\vector(1,0){70}}
\put(75,90){\makebox(0,0){\footnotesize Girard translation}}
\put(40,70){\vector(2,-1){70}}
\put(40,40){\makebox(0,0){\footnotesize CbN CPS translation}}

\put(180,5){\makebox(0,0){  traced closed Freyd category $\mathcal{C}\stackrel{F}{\rightarrow}\mathcal{D}$ }}
\put(180,-10){\makebox(0,0){  with $F(-)\otimes D\dashv D\Rightarrow(-)$}}

{ 
\put(180,15){\vector(0,1){85}} \put(177,15){\line(1,0){6}}
}

\put(180,120){\makebox(0,0){  compact closed category $\mathbf{Int}\,\mathcal{D}$}}
\put(180,110){\makebox(0,0){  with $!(X,U)=(U\Rightarrow X,1)$}}

{ 
\put(105,120){\vector(-1,0){35}} \put(105,117){\line(0,1){6}}
}

\put(-20,120){\makebox(0,0){  cartesian closed category of}}
\put(-20,110){\makebox(0,0){  finite products of cofree coalgebras of $!$}}

{ 
\put(90,0){\vector(-1,0){40}} \put(90,-3){\line(0,1){6}}
}

\put(-20,0){\makebox(0,0){  cartesian closed category of}}
\put(-20,-10){\makebox(0,0){  finite products of $D\Rightarrow E$'s}}

{ 
\put(-21,10){\line(0,1){90}} 
\put(-19,10){\line(0,1){90}} 
\put(-50,50){\makebox(0,0){equivalent}} 
}

\end{picture}

\end{center}

\section{Concluding remarks} \label{sec:concl}

We gave a translation from MELL to a lambda calculus with cyclic sharing.
This translation is derived from an easy theorem (Theorem \ref{thm:main})
on traced monoidal categories.
Once we know the theorem, it is fairly routine to derive the syntactic translation.
Perhaps the most difficult part would be to establish the appropriate syntax and 
categorical semantics of MELL and cyclic sharing, which had been sorted out 
many yeas ago. We do not claim that this translation would immediately
lead to a practical application, but hope that it makes an interesting case of the
semantics-driven approach to program transformations.

Our translation can be seen as a combination
of GoI interpretation and CPS translation.
Although we could have used any calculi/proof nets which are sound %and complete 
for $*$-autonomous categories with a linear exponential comonad,
the simple design of DCLL allows us to simplify the description of the translation a lot.
In particular, in this formulation with linear/non-linear implications,
the relation to the CPS translation is very easily observed. 
It seems that this relation to CPS semantics
is new; Sch{\"{o}}pp \cite{Sch14} observed coincidence of CPS semantics
and Int-interpretation in a different setting, but we are yet to see if there is
any formal relationship between his work and ours. It would also be meaningful
to investigate the connection between our translation and game semantics
(or tensorial logic)
along the work of Melli\`es and Tabareau \cite{MT09}, where
they study categorical structures closely related to ours.

Finally we shall mention a relation to the categorical
semantics (and game semantics) of the $\pi$-calculus in \cite{Lai05,ST17}.
The models in these work %\cite{Lai05,ST17} 
form traced closed Freyd categories, 
thus are instances of the 
structure considered in this paper. % (and motivated this work at its initial stage). 
It would be interesting to combine
our translation and their work, which might lead to new relation between
linear logic and concurrency theory.

\paragraph{Acknowledgements}
I thank Ken Sakayori and Takeshi Tsukada for stimulating discussions
related to this work.
This work was supported by 
JSPS KAKENHI Grant Numbers JP15K00013, JP18K11165
and
JST ERATO Grant Number JPMJER1603, Japan.

\nocite{*}
\bibliographystyle{eptcs}
\bibliography{letrec}

\end{document}